\documentclass[twocolumn]{article}
\usepackage[margin=0.6in,columnsep=0.35in]{geometry}
\usepackage{graphicx} 
\usepackage{xcolor}
\usepackage{tabularx}
\usepackage{amsmath}
\usepackage{amssymb}
\usepackage{subcaption}
\usepackage{multirow}
\usepackage{makecell}
\usepackage{url}
\usepackage{braket}
\usepackage[version=4]{mhchem}
\usepackage{placeins}
\usepackage{hyperref}
\usepackage{orcidlink}
\usepackage[numbers,sort&compress]{natbib}
\usepackage{caption}
\newcommand{\bigE}{\scalebox{1.8}{$\mathbb{E}$}}

\usepackage{authblk}

\usepackage{titlesec}

\titleformat{\section}
  {\normalfont\scshape\Large}
  {\thesection}{1em}{}
\titleformat{\subsection}
  {\normalfont\scshape\large}
  {\thesubsection}{1em}{}
\titleformat{\subsubsection}
  {\normalfont\scshape\normalsize}
  {\thesubsubsection}{1em}{}
  
\newif\ifarxiv
\arxivtrue

\begin{document}

\setlength{\parindent}{18pt}

\title{\scshape\Large Enabling ab initio geometry optimization of strongly correlated systems with transferable deep quantum Monte Carlo}
\author[1,$\dagger$]{P. B. Szab\'o\orcidlink{0000-0003-1824-8322}}
\author[1,$\dagger$]{Z. Sch\"{a}tzle\orcidlink{0000-0002-5345-6592}}
\author[1,2,3,4,*]{F. No\'{e}\orcidlink{0000-0003-4169-9324}}
\affil[1]{FU Berlin, Department of Mathematics and Computer Science,
Arnimallee 6, 14195 Berlin, Germany}
\affil[2]{Microsoft Research AI4Science, Karl-Liebknecht Str. 32, 10178 Berlin, Germany}
\affil[3]{FU Berlin, Department of Physics, Arnimallee 14, 14195 Berlin, Germany}
\affil[4]{Rice University, Department of Chemistry, Houston, TX 77005, USA}

\date{}

\twocolumn[{%
  \maketitle
  \vspace{-3em}
  \begin{center}
  \begin{minipage}{0.85\linewidth}
    \small

A faithful description of chemical processes requires exploring extended regions of the molecular potential energy surface (PES), which remains challenging for strongly correlated systems. 
Transferable deep-learning variational Monte Carlo (VMC) offers a promising route by efficiently solving the electronic Schrödinger equation jointly across molecular geometries at consistently high accuracy, yet its stochastic nature renders direct exploration of molecular configuration space nontrivial.
Here, we present a framework for highly accurate \textit{ab initio} exploration of PESs that combines transferable deep-learning VMC with a cost-effective estimation of energies, forces, and Hessians. 
By continuously sampling nuclear configurations during VMC optimization of electronic wave functions, we obtain transferable descriptions that achieve zero-shot chemical accuracy within chemically relevant distributions of molecular geometries. 
Throughout the subsequent characterization of molecular configuration space, the PES is evaluated only sparsely, with local approximations constructed by estimating VMC energies and forces at sampled geometries and aggregating the resulting noisy data using Gaussian process regression.
Our method enables accurate and efficient exploration of complex PES landscapes, including structure relaxation, transition-state searches, and minimum-energy pathways, for both ground and excited states. This opens the door to studying bond breaking, formation, and large structural rearrangements in systems with pronounced multi-reference character.

  \end{minipage}
  \end{center}
  \vspace{1em}
}]

\makeatletter
\def\blfootnote{\xdef\@thefnmark{}\@footnotetext}
\makeatother

\blfootnote{$^*$~Email: \href{mailto:frank.noe@fu-berlin.de}{frank.noe@fu-berlin.de}}%
\blfootnote{$^\dagger$~Equal contribution}%


The overarching objective of computational chemistry is to characterize chemical processes by exploring the molecular potential energy surface (PES).
This seemingly uniform prescription masks a great variety of problems arising due to the diversity of chemical processes themselves, comprising among others strongly correlated intermediates \cite{lischka2004,monino2022,dawes2016}, excited state processes \cite{ghosh2018,gonzalez2012}, or even light-matter interaction \cite{galego2019,tang2025}.
The wide range of scenarios in turn necessitated the development of a huge array of computational methods \cite{helgaker2013}, most of which can efficiently solve only a small subset of these problems.
Moreover, to make the electronic structure problem computationally tangible, the majority of quantum chemistry methods are designed to solve the electronic Schrödinger equation for a single molecular configuration at a time.
This necessitates intermediate steps to arrive from ab-initio electronic structure simulation to practical molecular modeling \cite{schlegel2003}, and renders the cost of directly exploring the PES with accurate quantum chemistry methods prohibitive for all but the smallest systems.
A common workaround is to use computationally less demanding methods to generate optimized molecular structures, then supply these geometries to more accurate methods to compute single-point energies.
While this is considered a workable solution for ground-state, single-reference equilibrium structures, where methods such as density functional theory (DFT) typically yield reasonable geometries, it can fail in several other scenarios such as those involving excited states \cite{bremond2018,wang2020} or strong static correlation \cite{minenkov2012}.
In such cases, the limitations of the cheaper approximate methods cast doubt on the quality of the resulting geometries, potentially rendering the subsequent energy evaluations with formally more accurate methods unreliable.
Another alternative is the optimization of surrogate models for predicting the PES from large numbers of single point simulations \cite{zhangGlobalInitioPotential2004,lu2020}.
This, however, requires careful construction datasets employing quantum chemistry methods tailored to specific regions of molecular configuration space \cite{boye-peronne2006}, which can be a costly and labor intensive process. 
Furthermore, working with surrogate models entails a loss of direct access to the \textit{ab initio} solution, and introduces additional sources of errors in form of intermediate computational steps.

\begin{figure*}[!t]
    \centering
 \makebox[\textwidth][c]{%
        \includegraphics[width=1\linewidth]{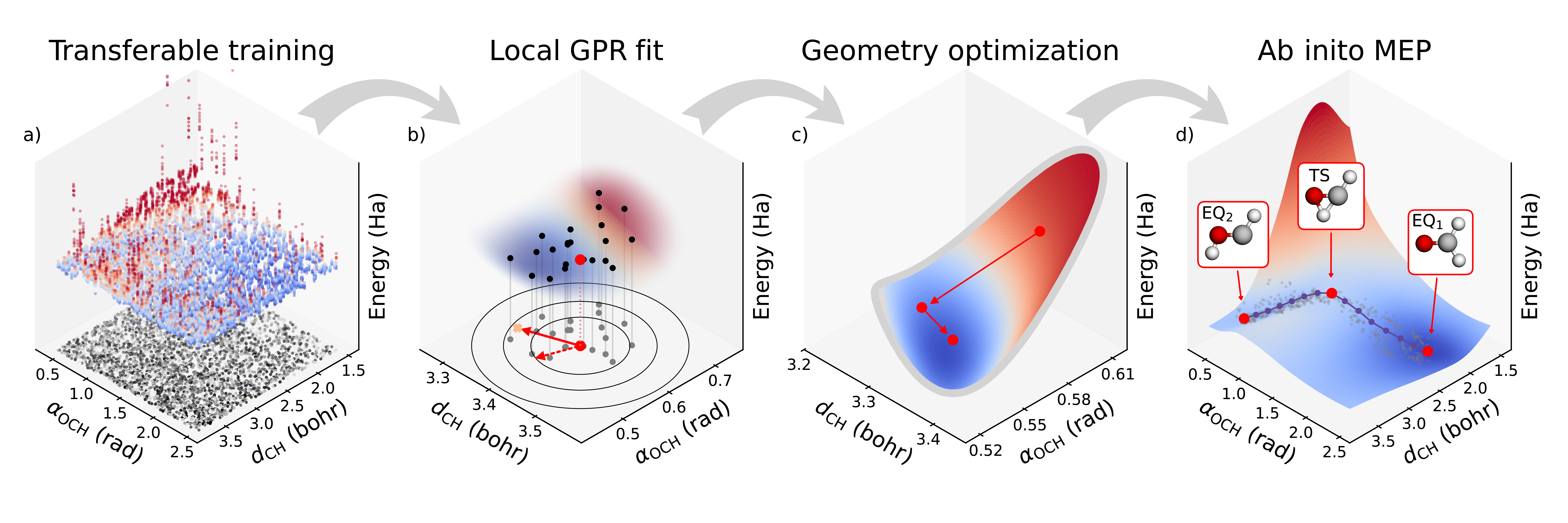}}
    \caption{
        \textbf{Sketch of the DeepQMC/GPR ab initio geometry optimization method.} The flowchart depicts the process of optimizing the minimum energy pathway (MEP) of the formaldehyde isomerization. a) A transferable wave function is optimized on continuously sampled nuclear configurations of a two dimensional cut of the formaldehyde configuration space. 
        Since the training loss consists of expectation values over small electron batches at random nuclear configurations, during transferable training the full PES is never materialized beyond noisy estimates. b) To extract energy and gradient information from the optimized transferable wave function, Monte Carlo estimators of the molecular energy and Hellmann-Feynman force are evaluated. The predictions are further improved by evaluating these expectation values at a range of molecular configurations in the vicinity of the geometry under investigation and fitting a local approximation of the PES with Gaussian process regression (GPR). From the Gaussian process energy, force and Hessian are computed analytically, which suffice to compute second order geometry updates. c) By a consecutive application of optimization steps, molecular geometries can be relaxed, transition states can be found and intermediates can be obtained when employing additional constrains. d) Combining multiple geometry optimizations, the full MEP of a chemical process can be characterized. Notably, for the MEP optimization the PES was only approximated sparsely via the samples depicted in gray.} 
    \label{fig:method}
\end{figure*}

In this work, we present a method that enables the \textit{ab initio} exploration of molecular PESs via variational Monte Carlo (VMC) optimization of transferable neural network wave functions.
In particular, we introduce a novel technique that exploits the generalization of transferable deep-learning ansatzes across molecular configuration space to efficiently evaluate the properties of the PES, including its value, gradient, and Hessian with respect to the nuclear coordinates.
Our electronic structure method of choice, deep-learning VMC, has been shown to provide state-of-the-art accuracy even for strongly correlated systems, while recent efforts towards a transferable ansatz have enabled the simultaneous modeling of a range of molecular configurations at a steeply discounted cost \cite{scherbela2022,gao2022,gao2022a,foster2025}.
Employing deep-learning VMC, we construct transferable wave function models that span broad regions of molecular configuration space at zero-shot chemical accuracy ($<$1~kcal/mol), trained at a cost comparable to that of a single-point simulation.
The universal applicability of deep-learning VMC allows us to account for strong correlation, as well as excited states, enabling the use of this single method for the consistent description of entire PESs as well as multiple electronic states \cite{schatzle2025}.
To enhance sampling efficiency and systematically account for the stochasticity of Monte Carlo simulation, we integrate deep-learning VMC with Gaussian process regression (GPR), fitted on the \textit{ab initio} energy and nuclear force estimates obtained from the transferable wave function.
This approach makes use of the generalization capability of transferable neural network ansatzes to unseen geometries, the zero-variance property of the energy, as well as the sound statistical formulation of GPR to deliver accurate estimates of the PES along with its first and second derivatives, complete with robust error measures.
While direct \textit{ab initio} geometry optimization with quantum Monte Carlo methods has been previously attempted and can yield accurate geometries for small molecules \cite{guareschi2013, barboriniStructuralOptimizationQuantum2012, iyerForceFreeIdentificationMinimumEnergy2024, kurian2026}, these methods are not matured, have not been applied using highly accurate deep-learning-based wave functions, and do not leverage the substantial efficiency gains offered by transferable wave function optimization, which we identify as a key innovation for achieving efficient and highly accurate geometry optimization in molecular systems with challenging electronic structure.
To demonstrate the resulting method's broad applicability, we perform a variety of geometry optimization tasks, such as geometry relaxation, transition state search, and minimum energy pathway (MEP) optimization on ground and excited state PESs for systems ranging from simple one dimensional diatomic molecules over small organic compounds to a nine dimensional scan of the \ce{HO2* + *OH} reaction.

\section{Results}\label{sec:results}
\subsection{Ab initio potential energy surfaces and geometry optimization with Gaussian process regression}\label{sec:results:method}

\textit{Ab initio} geometry optimization involves the successive solution of the electronic Schrödinger equation to evaluate derivatives of the nuclear potential energy, thereby guiding the system toward extrema on the Born--Oppenheimer PES.
Our method provides an effective solution to this problem that leverages the excellent accuracy of deep-learning VMC, combining the transferable simulation of electronic states with an appropriate treatment of the stochasticity inherent to Monte Carlo estimation of expectation values. The first step of the proposed algorithm is the transferable optimization of a neural network wave function ansatz $\Psi_{\boldsymbol{\theta}}(\mathbf{r}|\mathbf{R})$, aiming at zero-shot chemical accuracy on large, continuous, and chemically relevant regions of the PES.
To this end, we continuously sample molecular configurations during wave function optimization
\begin{equation}
    \boldsymbol{\theta}^* = \underset{\boldsymbol{\theta}}{\operatorname{argmin}}~\underset{\mathbf{R}\sim \rho(\mathbf R)}{\bigE}~\underset{{\mathbf{r}\sim |\Psi_{\boldsymbol{\theta}}(\mathbf{r}\left|\mathbf{R})\right|^2}}{\bigE}\left[ \frac{\hat{H}(\mathbf R)\Psi_{\boldsymbol{\theta}}(\mathbf{r}|\mathbf{R})}{\Psi_{\boldsymbol{\theta}}(\mathbf{r}|\mathbf{R})}\right]\,,
\end{equation}
where $\hat{H}(\mathbf{R})$ is the electronic Hamiltonian with the external potential of a set of nuclei in configuration $\mathbf{R}$, and $\rho(\mathbf{R}): \Omega\rightarrow\mathbb{R}_{\geq0}$ is a probability density over the relevant subspace of the molecular configuration space $\Omega\subseteq\mathbb{R}^{3M}$, i.e. a uniform distribution over bond lengths, angles, and torsions.
While nuclear geometries of the outer expectation values are sampled directly, the expectation value over the electron coordinates $\mathbf r$ depends on the molecular geometry through the transferable wave function and is evaluated via Markov chain Monte Carlo.
To maintain high sampling efficiency, we employ a space-warp technique to update electron coordinates in tandem with changes in the molecular geometry \cite{umrigar1989}.
This continuous transferable optimization yields wave functions with consistent energies and forces throughout the range of configurations at an overall training cost comparable to an individual single-point optimization.
A study that establishes the relevance of continuous sampling for consistent zero-shot accuracy and demonstrates the computational benefits of the transferable simulation is presented in Sec.~A.1 of the Supplementary Information. 

With the optimized transferable wave function at hand we turn to the exploration of PESs.
Working with the wave function directly means that during inference, expectation values of the energy and force have to be estimated with Monte Carlo integration.
The stochasticity of Monte Carlo estimation is a known challenge when exploring PESs \cite{iyerForceFreeIdentificationMinimumEnergy2024, gao2022a}, that we account for by employing advanced zero-variance force estimators \cite{assaraf1999,assaraf2003}, and by using the \textit{ab initio} energy and force data to fit statistically optimal local PES models using GPR \cite{bartok2010, deringer2021}.
GPR for molecular PESs fits non-parametric models to make energy predictions for unseen molecular geometries based on observations.
Gaussian processes (GPs) provide uncertainty measures for their predictions and facilitate the computation of analytic gradient and Hessian estimates.
We acquire a local approximation of the PES, by conditioning a GP on datasets $\mathcal{D}$ of energies and forces $\mathbf{Y} = [\bar{E_1}, \bar{\mathbf{F}}_1, \ldots, \bar{E}_n, \bar{\mathbf{F}}_n]^\mathrm{T}$ at molecular configurations $\mathbf{X} = [\mathbf{R}_1, \ldots, \mathbf{R}_n]^\mathrm{T}$, sampled within a neighborhood of the geometry of interest $\mathbf{R}$. The estimates $\bar{E}_i$ and $\bar{\mathbf{F}}_i
$ are obtained by evaluating the appropriate Monte Carlo estimators on the optimized transferable wave function $\Psi_{\boldsymbol{\theta}^*}$
\ifarxiv
\begin{align}
\begin{split}
    \big(\bar{E}_i,\bar{\mathbf{F}}_i\big) &= \Big(\big\langle \hat H(\mathbf{R}_i)\big\rangle_{\Psi_{\boldsymbol{\theta}^*}(\mathbf{r}|\mathbf{R}_i)}\raisebox{-1.1ex}{,} \big\langle\left.{\mathrm{d}\hat H}/{\mathrm{d} \mathbf{R}}\right|_{\mathbf{R}_i} \big\rangle_{\Psi_{\boldsymbol{\theta}^*}(\mathbf{r}|\mathbf{R}_i)}
    \Big)\\ &\approx \Big(E(\mathbf{R}_i),\mathbf{F}(\mathbf{R}_i)\Big)\,.
\end{split}
\end{align}
\fi
Conditioning the GP on the data yields a random variable for the energy
\begin{equation}
    \tilde{E}(\mathbf{R})  \mid \mathcal{D}\sim\mathcal{N}(m^*(\mathbf{R}),k^*(\mathbf{R},\mathbf{R}))\,,
\end{equation}
where the optimal posterior mean $m^*$ and posterior variance $k^*$ can be obtained in closed form and constitute the prediction of the energy and its uncertainty, respectively.
An important property of GPs is that their derivatives are themselves GPs, enabling the direct prediction of forces and Hessians. 

Throughout the geometry optimization, at each new molecular configuration we query the GP fitted on the previous energy and force estimates at negligible cost, and check whether it has an error estimate below a set accuracy threshold.
While this is not the case, we sample nuclear configurations in the vicinity of the geometry of interest to obtain additional data points until sufficient accuracy of the GPR is achieved. 
The energy, nuclear force, and Hessian information obtained from the GP is then used for downstream PES exploration tasks, such as geometry relaxation or minimum energy pathway (MEP) search. 
As demonstrated in Sec.~A.2 of the Supplementary Information, our GP force estimates are at least as accurate and efficient as the best zero-bias zero-variance force estimators even if only a single geometry is of interest, with further efficiency benefits arising in the geometry optimization setting from reusing the data generated previously for other nearby geometries.
Since the dataset for GPR is constructed on-the-fly, materializing the full PES can be avoided.

With accurate local information about the PES at hand, geometry optimization is carried out using standard techniques, in particular via the Newton--Raphson method in internal coordinates \cite{bakken2002}, employing image functions for transition state searches \cite{helgaker1991}, and the reduced-restricted approach of Anglada for constrained optimization \cite{anglada1997,devico2005}.

\subsection{Validating equilibrium structures of diatomic molecules}
\ifarxiv
\begin{figure}[h]
\centering
    \includegraphics[width=\linewidth]{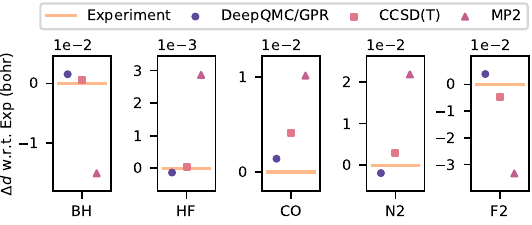}
   \caption{\textbf{Optimizing equilibrium bond lengths for diatomic molecules.} DeepQMC/GPR results are compared to experimental data \cite{huber1979} and to MP2@CBS and CCSD(T)@cc-pV6Z calculations \cite{pawlowski2003}.}
   \label{fig:diatomics}
\end{figure}
\fi

\begin{figure*}[!t]
    \centering
    \includegraphics[width=\linewidth]{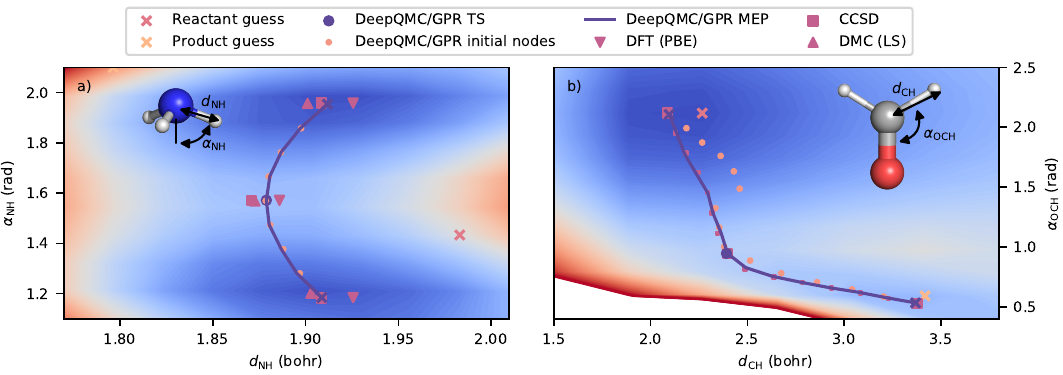}
    \caption{
        \textbf{Minimum energy path search on two dimensional cuts of the ammonia and formaldehyde potential energy surfaces.} On panel a) the minimum energy pathway for the inversion of ammonia is shown, while panel b) depicts the isomerization of formaldehyde.
        Reference and baseline results for the ammonia inversion are taken from Ref. \cite{iyerForceFreeIdentificationMinimumEnergy2024}, CCSD results for formaldehyde are computed in house. The heatmaps visualize PESs extracted from the transferable deep-learning VMC solution and serve as a guide to the eye.}
    \label{fig:nh3-formaldehyde-mep}
\end{figure*}
Before extending to higher dimensional geometry optimizations, we validate our method by comparing equilibrium bond lengths of diatomic molecules optimized with transferable deep-learning VMC to experimental and computational references.
We follow the procedure outlined in Sec.~\ref{sec:results:method} for a test set of five small molecules, namely BH, HF, CO, N\textsubscript{2}, F\textsubscript{2}.
Throughout transferable training, bond lengths are sampled from a mixture of four Gaussian distributions with $\mu_0=1.5\,\text{bohr},~\mu_1=2.5\,\text{bohr},~\mu_2=3.5\,\text{bohr}$ and $\mu_3=4.5\,\text{bohr}$ and $\sigma=1\,\text{bohr}$.
To avoid high energy configurations we restrict the interatomic distance to be larger than $1\, \text{bohr}$ for HF and F\textsubscript{2} and $1.5\,\text{bohr}$ for BH, CO and N\textsubscript{2}, respectively.
Using the converged transferable wave functions, geometry optimizations from a stretched bond length of $3.5\,\text{bohr}$ are commenced. 
The relaxations converge within approximately ten iterations to a mean absolute error (MAE) with respect to the experimental reference \cite{huber1979}  of $1.8\times10^{-3}\,\text{bohr}$.
This is similar to the MAE of $2.5\times10^{-3}\,\text{bohr}$ of CCSD(T) calculations in the large cc-pV6Z basis and significantly lower than the MAE of $1.6\times10^{-2}\,\text{bohr}$ MP2 calculation extrapolated to the complete basis set limit \cite{pawlowski2003}.
The close agreement with experimental references and gold-standard CCSD(T) geometry optimizations for equilibrium structures, which are typically of single-reference character, supports the reliability of our method.

\subsection{Exploring reactivity with minimum energy path searches}

After validating the accuracy of our approach on diatomics, we now turn to showcasing the methodology on the significantly more complex task of MEP optimizations.
Being the lowest energy pathways on the PES that connect two local minima (reactants and products), MEPs are a central concept in the theoretical description of reactivity.
Given their relevance, several approaches have been devised for their computation, such as the growing string \cite{peters2004} and nudged elastic band \cite{mills1994} methods, all of which operate by iteratively refining a chain of geometries along the pathway.
As constructing each link of the chain requires several geometry optimization steps, the computation of longer chains can necessitate dozens or hundreds of PES evaluations. In this setting, properly exploiting the capabilities of transferable deep-learning VMC can result in an enormous computational advantage, which could be the key to unlocking routinely obtainable, highly accurate, \textit{ab initio} MEPs for strongly correlated systems.

To test the performance of the proposed transferable PES estimation approach, we implement a simple version of the double-ended growing string method \cite{peters2004}.
Described in more detail in Sec.~\ref{sec:methods:mepopt}, this algorithm starts by optimizing the structures of two local minima (reactant and product), and proceeds to connect them via a chain of images.
Then, a transition state search is carried out starting from the image with the highest energy, after which the original chain is replaced with new strings connecting the transition state with the reactant and the product.

The algorithm is first applied to the inversion of the ammonia molecule, where traditional VMC, diffusion Monte Carlo (DMC), CCSD(T) and DFT references are available \cite{iyerForceFreeIdentificationMinimumEnergy2024}.
Figure \ref{fig:nh3-formaldehyde-mep}a), displays a two-dimension cut of the ammonia PES, with all nitrogen-hydrogen bonds constrained to be the same length, and at the same angle to the rotational symmetry axis of the molecule.
A transferable neural network ansatz was trained for continuously sampled geometries in the range $1.77 ~\text{bohr} \le d_{\ce{NH}} \le 2.00 ~\text{bohr}$ and $\frac{\pi}{3} \le \alpha_{\ce{NH-z}} \le \frac{2\pi}{3}$. 
With a total number of 200k training iterations and a total electron batch size of 4096 the overall cost of the transferable optimization was equivalent to a small number of standard single point simulations.
The trained ansatz was then used to optimize the structures of the two local minima, both of which are in excellent agreement with the reference CCSD geometries.
Lastly, the MEP is constructed as a chain of seven intermediate images, from which the middle one represents the transition state, lying neatly among the reference transition state geometries, and yielding a reaction barrier height of 5.3 kcal/mol, that matches well with the reference CCSD, CCSD(T), and DMC values of 5.6, 5.8, and $4.4\pm 0.7$ kcal/mol, respectively.

\begin{figure*}[!t]
    \centering
    \includegraphics[width=\linewidth]{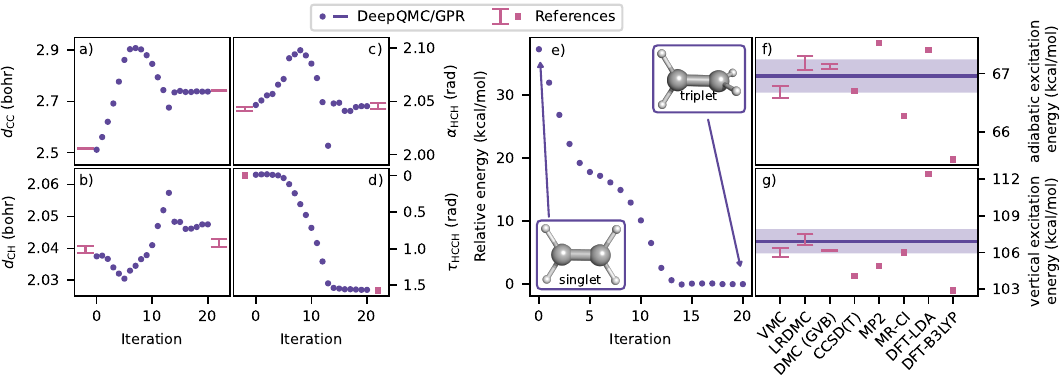}
    \caption{\textbf{Adiabatic excitation energies for the ethylene molecule.} On panels a)-d) the internal coordinates of the relaxation trajectory of the triplet state after a vertical excitation from the singlet ground state equilibrium geometry are depicted. The VMC \cite{barboriniStructuralOptimizationQuantum2012} reference equilibrium geometries of the singlet and triplet states are well reproduced. Corresponding estimates of the relative energy with respect to the triplet equilibrium geometry are displayed on panel e). The inserts show the optimized equilibrium geometry of the singlet and triplet state. Panel f) and panel g) compare the adiabatic and the vertical excitation energies to reference calculations \cite{barboriniStructuralOptimizationQuantum2012}, where the shaded area indicates chemical accuracy.}
    \label{fig:ethylene}
\end{figure*}

Next, an MEP search is carried out for the slightly more involved process of formaldehyde isomerization, where one of the hydrogen atoms migrates from the carbon to the oxygen atom, breaking the old and forming a new covalent bond along the way.
As before, a single neural network ansatz is trained for the two dimensional PES depicted on Figure \ref{fig:nh3-formaldehyde-mep}b) and defined by $1.5 ~\text{bohr} \le d_{\ce{CH}} \le 3.8 ~\text{bohr}$, and $0.4 \le \alpha_{\ce{OCH}} \le 2.5$.
The corresponding MEP comprises a narrow channel near the product and a turn at the transition state.
Accordingly, more intermediate images are used here, which further increases the required number of PES evaluations.
The deep-learning VMC MEP is compared with a reference CCSD path computed in house.
It is clear that all intermediate geometries, as well as the transition state agree quite well between the two methods, while the mean absolute deviation of relative energies across the pathway is as low as 0.5 kcal/mol.
In the process of the full MEP optimization a total of 133 geometry updates were performed, and the fact that a single \textit{ab initio} deep-learning VMC data point was used in determining 13.8 updates on average highlights the data efficiency of the GP based force estimator on realistic tasks.

\subsection{Geometry optimization of electronic excited states}
\label{sec:results:ethene}
\begin{figure*}[!t]
    \centering
    \includegraphics[width=\linewidth]{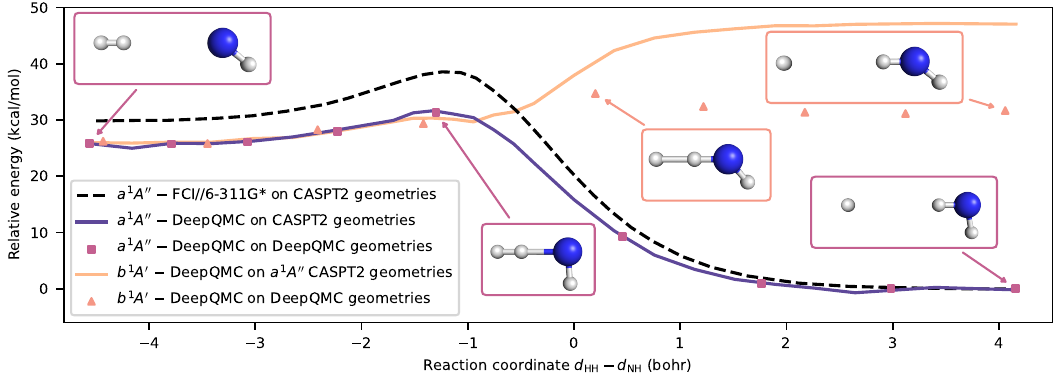}
    \caption{\textbf{Potential energy surfaces of the \ce{H2 + NH -> H* + *NH2} reaction.} The reaction might proceed either on the $a^1A''$ or the $b^1A'$ PES. Inset plots show the relevant reactant, transition state, and product structures highlighting the differences between the two electronic states. The FCI results taken from Wu \textit{et al.} \cite{wu2020} were computed on geometries optimized at the CASPT2 level of theory, with the relatively small basis set of 6-311G*. The energies of these geometries were also evaluated with DeepQMC, shown with solid lines.}
    \label{fig:hnh2}
\end{figure*}
The computation of PESs via transferable deep-learning VMC can be straightforwardly extended to electronic excited states \cite{schatzle2025}.
In this section, we exploit this general applicability, to demonstrate that the approach presented here is capable of accurately determining adiabatic excitation energies, as well as studying the reactivity of molecules in their excited states.

To this end, we first study ethylene, as a common model for photochemistry that undergoes large structural changes upon singlet-triplet excitation.
We train a transferable neural network ansatz for the singlet ground state and first triplet excited state of the ethylene molecule employing the spin-penalty method introduced in \cite{szabo2024}.
The setup of \citet{barboriniStructuralOptimizationQuantum2012} is adopted, with the molecule being parameterized in terms of the \ce{H-C-C-H} torsion angle, the \ce{H-C-H} angle, the \ce{C-H} bond length, and the length of the \ce{C-C} double bond, forcing the same values for all occurrences of each. 
This yields a four dimensional configuration space for the transferable ansatz and geometry optimization.
Utilizing the transferable wave function for the singlet state, we first relax the geometry of the ground state to it's global minimum.
Next, ethylene's vertical excitation energy is computed by evaluating the energy of the triplet state at the ground state equilibrium geometry.
As displayed on panel g) of Figure \ref{fig:ethylene}, our vertical excitation energy estimate is in excellent agreement with other QMC derived values, as well as multi-reference CI, while the single reference methods CCSD(T) and MP2 exhibit slightly larger deviations and DFT estimates vary significantly based on the functional being used \cite{barboriniStructuralOptimizationQuantum2012}.
Lastly, the ground state minimum geometry is relaxed on the triplet surface in order to compute the adiabatic gap.
The four dimensional optimization trajectory is depicted in panels a)-d) of Figure \ref{fig:ethylene}, with the relative energy with respect to the optimized triplet equilibrium geometry depicted on panel e).
Our simulations reproduce the 90$^\circ$ \ce{H-C-C-H} torsion and the elongation of the \ce{C-C} double bond of the triplet minimum, leading to an energy reduction of 36 kcal/mol. 
The adiabatic excitation energy is displayed on panel f).
Our results are again in excellent agreement with the reference QMC values being within chemical accuracy of our estimate.
This indicates that the proposed geometry optimization procedure successfully translates the accuracy of deep-learning VMC to relative energies between different geometries and electronic states.

With applicability established to adiabatic excitations, we move to the task of characterizing elementary chemical reactions proceeding entirely on excited surfaces.
The radical-radical hydrogen abstraction reaction of \ce{H2 + NH -> H* + *NH2} is selected as our next target.
In a recent study \cite{wu2020}, the possibility of this process proceeding on various electronic states was discussed.
Commencing on the ground state triplet surface, the reaction is well described by single reference methods.
On the other hand, the reaction might start from either of the low-lying, doubly-degenerate singlet excited states of \ce{NH}, leading to products either in the $b^1A'$ or the $a^1A''$ electronic state, having the unpaired electron of \ce{NH2} either in or out of the molecule's plane, respectively.
To investigate reactivity in these states, a transferable neural network ansatz is trained for the $b^1A'$ and $a^1A''$ surfaces, considering only planar geometries and a collinear attack of the \ce{H2} molecule (see Sec.~B.2 of the Supplementary Information for the definition of the training geometry distribution).
In the work of Wu \textit{et al.}, geometries along the reaction pathway on the $a^1A''$ surface are optimized at the CASPT2@6-311G* level of theory, while the reaction energy and barrier height of the reverse reaction are determined to be 29.9 and 38.6 kcal/mol, respectively, via full configuration interaction (FCI) calculations in the same basis, as shown on Figure~\ref{fig:hnh2}.
Evaluating the energies of the CASPT2 geometries using our neural network ansatz, we obtain estimates of the reaction energy and barrier height that are about 4.0 and 7.3 kcal/mol lower than those obtained with FCI, highlighting the importance of finite basis set errors of second quantized methods, when using insufficiently large basis sets.
Next, to compare reactivity between the two surfaces, the deep-learning VMC energies of the CASPT2 geometries are also evaluated on the $b^1A'$ surface, as plotted on Figure \ref{fig:hnh2}, resulting in a practically barrierless trajectory that increases in energy towards the products.
Finally, performing our full MEP search on the $b^1A'$ surface, we find that the \ce{NH2} radical relaxes by 15.4 kcal/mol into a configuration with the H--N--H angle increased to 139$^\circ$.
As displayed on Figure \ref{fig:hnh2}, on the relaxed $b^1A'$ pathway we obtain for the reverse reaction a reaction energy of $-5.5$ kcal/mol and a barrier height of 3.0 kcal/mol, the latter being significantly lower than in the $a^1A''$ state, indicating that the reverse reaction might take place much more readily on this excited $b^1A'$ surface.

Taken together, the results presented in this section show that the \textit{ab initio} geometry-optimization protocol based on transferable deep-learning VMC facilitates the study of excited-state PESs, offering a powerful new tool for the investigation of processes involving photoexcitations of strongly correlated systems.

\subsection{Full potential energy surface for \ce{HO2 + OH} reaction}
\begin{figure*}[h!]
    \centering
    \includegraphics[width=\linewidth]{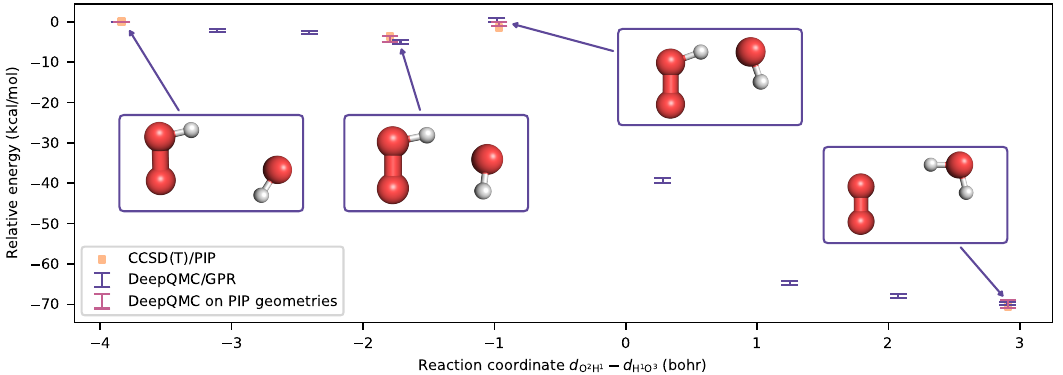}
    \caption{{\bf MEP for the \ce{HO2* + OH* -> H2O + O2} reaction.} The results of the deep-learning VMC MEP optimization are compared with reference values obtained with a permutation invariant polynomial (PIP) neural network potential trained on 108k CCSD(T)-F12 single-point energies \cite{liuAnomalousKineticsReaction2019}.}
    \label{fig:h2o3-mep32-curve}
\end{figure*}

Lastly, we showcase the scalability of the proposed methodology, targeting the practically relevant problem of characterizing the radical-radical reaction \ce{HO2* + OH* -> O2 + H2O}.
This process is of high importance for both atmospheric and combustion chemistry, as a major consumer of \ce{OH*} and \ce{HO2*} radicals \cite{heard2003,hong2010}, while at the same time being difficult to characterize experimentally owing largely to the problematic determination of \ce{HO2*} radical concentrations as well as its radical-radical nature \cite{burke2013}.
Recently, a full PES has been computed for this process at the explicitly correlated CCSD(T)-F12 level of theory, by fitting a permutation invariant polynomial (PIP) neural network to more than hundred thousand coupled cluster energies \cite{liuAnomalousKineticsReaction2019}.
The authors emphasize the multi-reference nature of the system in question, especially in the entrance channel up to the transition state, due to the near-degeneracy of singlet and triplet spin states.
Nonetheless, this single-reference method yields results that are in agreement with some experiments.
Here, we aim to demonstrate that our proposed methodology is a useful tool to tackle such a realistic chemical problem, where the application of traditional single-reference electronic structure methods requires significant computational resources, while not dispelling all doubts regarding strong correlation effects. 

To this end, a transferable neural network ansatz is optimized on a nine dimensional region of the configuration space around the MEP identified by Liu \textit{et al.} \cite{liuAnomalousKineticsReaction2019} (for the procedure used to obtain this configuration space see Sec.~B.3 of the Supplementary Information).
During this task, sixteen thousand distinct geometries were visited, constituting the largest single molecule configuration space, for which a transferable neural network ansatz yields accurate, zero-shot predictions.
The pathway obtained with this transferable ansatz and our MEP finding algorithm is compared with the results reported by \citet{liuAnomalousKineticsReaction2019} on Figure \ref{fig:h2o3-mep32-curve}, with numerical values collected in Table \ref{tab:h2o3_data}.
Considering first the reactant and the product, we observe excellent agreement between CCSD(T) and deep-learning VMC, both in terms of molecular geometry (lower half of Table \ref{tab:h2o3_data}), as well as relative energy (upper half of Table \ref{tab:h2o3_data}).
This is in line with the facts that all degrees of freedom within the reactant and product fragments are relatively stiff, and that both CCSD(T) and deep-learning VMC are expected to give accurate descriptions of such single reference structures.
\begin{table*}[!ht]
    \centering
    \caption{{\bf Comparison of the deep-learning VMC and CCSD(T) characterizations} of the \ce{HO2* + OH* -> H2O + O2} reaction. TS denotes the transition state structure, while CP1 corresponds to the local minimum found near the entrance of the reactive channel, first described in Liu \textit{et al.} \cite{liuAnomalousKineticsReaction2019}. Two sets of geometries are considered: those obtained from the PIP neural-network potential fitted to CCSD(T)-F12a energies, and those optimized via the presented DeepQMC/GPR method.
    }
    \begin{tabularx}{\textwidth}{XX|cccc}
    &                     & reactant & CP1    & TS           & product  \\ \hline
        \multirow{3}{*}{\makecell{relative energy \\ (kcal/mol)}}  & CCSD(T)/PIP         & $0.0$    & $-3.7$ & $-1.4$       & $-70.6$  \\
                                                                   & DeepQMC/GPR         & $0.0$    & $-4.9\pm0.5$ & $0.4\pm0.5$       & $-69.8\pm0.4$  \\
                                                                   & DeepQMC on PIP geom & $0.0$    & $-4.3\pm0.6$ & $-0.5\pm0.6$ & $-69.9\pm1.0$  \\ \hline
        \multirow{3}{*}{\makecell{DeepQMC v. PIP \\ geometry MAE}} & bonds (a$_0$)       & $0.02$   & $0.03$ & $0.01$       & $< 0.01$ \\
                                                                   & angles (deg)        & $0.7$    & $7.7$  & $4.3$        & $< 0.1$  \\
                                                                   & torsions (deg)      & ---      & $12$   & $3$          & ---      \\
    \end{tabularx}
    \label{tab:h2o3_data}
\end{table*}

Moving to the intermediate geometries, in particular to the local minimum situated in the entrance of the reaction channel (CP1), we find larger deviations in molecular structure between coupled cluster and DeepQMC, especially in angles and torsions.
To disentangle these differences, we also evaluate our transferable neural network ansatz on the geometries obtained via the CCSD(T)/PIP PESs (second row of Table \ref{tab:h2o3_data}).
The two sets of deep-learning VMC relative energies agree with each other within the reported single $\sigma$ error bounds, indicating that the differences between the CCSD(T) and deep-learning VMC geometries have limited energetic relevance.
Indeed, the largest contributors to these geometric differences are the inter-fragment (\ce{HO2} -- \ce{OH}) coordinates, which are expected to be the least stiff in the CP1 configuration.
For a stochastic method like deep-learning VMC, such a relatively flat PES results in a larger variance in the optimized relative orientation of the two fragments.
The same, to a lesser extent, is true for the transition state structure (TS), where we also find non-negligible differences (although noticably smaller than for CP1) in the inter-fragment angles and torsions between deep-learning VMC and CCSD(T).
Turning now to the relative energies of these intermediate geometries, we find that deep-learning VMC consistently stabilizes CP1 and destabilizes the transition state compared to CCSD(T), regardless of the exact geometry used.
It is therefore conceivable that this effect is a manifestation of the stronger static correlation to be expected for these geometries, and highlights the benefits of using black-box methods applicable to strongly correlated systems such as deep-learning VMC when describing PESs of chemical reactions.


\section{Summary and Discussion}
We have presented a novel method that enables the exploration of \textit{ab initio} ground and excited state potential energy surfaces of strongly correlated systems by integrating continuous transferable deep-learning VMC simulation with GPR.
Our method turns the extrapolation capabilities of transferable neural network wave functions into a practical approach for simulating chemical processes, amortizing the cost of transferable neural network wave function optimization through repeated evaluation across molecular configuration space.
This work provides the first demonstration of transferable wave functions achieving zero-shot chemical accuracy in relative energies across large, continuous regions of molecular configuration space, exemplified for systems with up to nine degrees of freedom.
Utilizing these ansatzes, advanced Monte Carlo estimators are employed to obtain low variance estimates of energy and nuclear force expectation values.
This data is then used to fit local GPR models of the PES, without sacrificing accuracy in the single-point setting, while greatly increasing data efficiency during geometry optimization, where the PES must be evaluated at a number of nearby points.

We illustrate the accuracy and broad usability of our method by targeting a range of distinct applications.
We initiate the assessment by relaxing five diatomic molecules, reaching very good agreement with experimental references.
Turning to the more demanding task of characterizing elementary chemical processes, we obtain results matching reference calculations when simulating the MEPs on the two dimensional PESs of the ammonia inversion and formaldehyde isomerization.
Next, the feasibility of accurate geometry optimizations on excited PESs is established, by computing the adiabatic excitation energy of the singlet-triplet excitation of ethylene, in agreement with computational references on both excitation energy, and excited state equilibrium geometry. 
We then proceed to characterize the \ce{H2 + NH -> H* + *NH2} reaction, qualitatively reproducing the small-basis FCI energy profile of the reaction proceeding in its lowest singlet state, and extending the analysis to the first singlet excited state, where a new analogous reaction channel is identified.
Lastly, we simulate a nine dimensional PES section around the \ce{HO2* + OH* -> H2O + O2} reaction pathway, demonstrating the scalability of our method to unprecedented volumes of molecular configuration space.
Using a single transferable ansatz for the entire PES, quantitative agreement is reached with a reference neural-network potential fitted on more than one hundred thousand CCSD(T)-F12a single-point energy calculations.
Small deviations around the intermediate geometries indicate that the use of multi-reference algorithms such as deep-learning VMC might be warranted.

Combining transferable QMC and GPR for the optimization of molecular geometries is a novel paradigm that leverages the interpolation capabilities of neural networks while retaining the computation's \textit{ab initio} character.
Anchored in deep-learning VMC, the method inherits all its beneficial attributes, such as its applicability to strongly correlated systems, a consistent description of all molecular configurations, i.e. the intermediates of chemical reactions, the straight forward extension to the simulation of excited states, and its overall black-box nature.
Furthermore, the method provides direct access to the wave function, and hence all electronic properties, at any point in configuration space and will benefit from any improvements made in the rapidly developing field of transferable deep-learning VMC.
While not strictly necessary, GPR further improves efficiency of our method by reusing information from previous force and energy evaluations of the PES and enables the automation of geometry optimizations by querying the costly deep-learning VMC oracle only when necessary.
Overall our method is a tool for the exploration of intricate PESs of small molecules that streamlines the process of quantum chemistry computation, fitting the PES with surrogate models and the optimization of chemical processes and significantly reduces the cost compared to the alternative of consecutive, independent \textit{ab initio} simulations.

While the presented results are promising, important challenges remain.
We estimate the simulation of the nine dimensional PES of \ce{H2O + O2} to be at the limit of our current capabilities. 
To increase system sizes in terms of the number of electrons as well as the dimensionality of the molecular configuration space technical innovations, such as the integration with finite range embeddings \cite{scherbela2025} or sparse wave function architectures \cite{li2024}, will be required. 
Furthermore, the method would greatly benefit from more mature and advanced geometry optimization protocols and could be employed as a tool for \textit{ab initio} molecular dynamics simulation.
We hope that our method will be further developed, and can become the default scheme for the PES exploration of strongly correlated, medium size systems.

\section{Methods}
The method presented here is concerned with solving two main tasks: obtaining neural network ansatzes that accurately describe the electronic structure for a large and continuous range of molecular geometries, and the subsequent extraction of all information necessary for geometry optimization from these ansatzes.
The rest of this section details the techniques used to accomplish these two objectives, and is organized as follows.
In Section \ref{sec:methods:qmc}, we describe solving the Schr\"odinger equation for a continuous range of molecular geometries with transferable deep-learning VMC.
Then, in Section \ref{sec:methods:nuclear_forces}, the techniques of obtaining energies, nuclear forces and Hessians from converged deep-learning VMC solutions are discussed, along with some aspects of utilizing this information in downstream tasks such as geometry or MEP optimizations in Section \ref{sec:methods:geomopt}.

\subsection{Transferable deep-learning variational Monte Carlo for continuous molecular configuration spaces}\label{sec:methods:qmc}
This section discusses the core concepts of deep-learning variational Monte Carlo as well as its extension to the geometrically transferable setting.
The generalization to continuous configuration spaces is comprised three main contributions, the transferable optimization across molecular geometries, the extended wavefunction ansatz and the sampling of electronic and nuclear degrees of freedom.

\subsubsection{Deep-learning variational Monte Carlo}
The time-independent, non-relativistic Schr\"odinger equation takes the form of an eigenvalue equation
\begin{equation}
    \hat H \Psi = E \Psi ~~~ \Psi \in \mathcal{H}^- \, ,
\end{equation}
where $\hat H$ is the Hamiltonian of the physical system under investigation, $\Psi$ is the wave function which, describing fermions, must belong to the anti-symmetric subspace of the many-body Hilber-space ($\mathcal{H}^-$) and $E$ is its associated energy.
Here, we are concerned with the behavior of electrons in molecular systems subject to the Born--Oppenheimer approximation, and therefore with the clamped nucleus Coulomb  Hamiltonian (in atomic units)
\begin{equation}
    \hat H = 
        -\frac{1}{2}\sum_i\nabla_i^2
        + \sum_{i<j} \frac{1}{r_{ij}}
        - \sum_{iI}\frac{Z_I}{r_{iI}} 
        {\color{gray}+ \Big(\sum_{I<J} \frac{Z_I Z_J}{r_{IJ}}\Big)}\, ,
\end{equation}
with indices $i$ and $j$ running over electrons, $I$ and $J$ over nuclei, $Z$ denoting nuclear charges, and $r$ interparticle distances.
Since electrons are considered in the static potential of the nuclei, the last term in the molecular Hamiltonian constitutes an energy offset.

Variational methods such as deep-learning VMC rely on the variational principle of quantum mechanics to approximate the solutions of the above equation via minimizing the Rayleigh quotient
\ifarxiv
\begin{align}
    \begin{split}
         \mathcal L^\text{sp}\big[\Psi_{\boldsymbol{\theta}}\big] &= \frac{\int d\mathbf{r}\Psi_{\boldsymbol{\theta}}^*(\mathbf{r})\hat{H}\Psi_{\boldsymbol{\theta}}(\mathbf{r})}{\int d\mathbf{r}|\Psi_{\boldsymbol{\theta}}(\mathbf{r})|^2}\\&=~\underset{{\mathbf{r}\sim |\Psi_{\boldsymbol{\theta}}(\mathbf{r}\left|\mathbf{R})\right|^2}}{\bigE}\left[ \frac{\hat{H}\Psi_{\boldsymbol{\theta}}(\mathbf{r})}{\Psi_{\boldsymbol{\theta}}(\mathbf{r})}\right]
    \label{eqn:rayleigh}
    \end{split}
\end{align}
   \fi
with respect to the parameters $\boldsymbol \theta$ of a trial wave function $\Psi_{\boldsymbol \theta}$.
For the molecular Hamiltonian, the wave function can be chosen real-valued $\Psi: \mathbb{R}^{3N}\rightarrow \mathbb{R}$ and we employ spin-assigned wave functions \cite{foulkes2001, schatzle2023}.
In variational Monte Carlo the high dimensional integrals in \eqref{eqn:rayleigh} are approximated using Markov chain Monte Carlo (MCMC) with importance sampling.
In practice, iterative minimization is employed where in each iteration the expectation value is estimated on a finite batch of samples drawn from the current wave function $p \sim |\Psi_{\boldsymbol \theta}(\mathbf{r})|^2$, which is utilized to compute a stochastic approximation of the energy gradient for updating the ansatz's parameters.
The flexibility of the Monte Carlo integration scheme enables the use of highly expressive ansatzes parametrized by neural networks, which is the key innovation behind the recent surge in accuracy of VMC methods.

Working in the Born--Oppenheimer approximation, we are concerned with solving the Schrödinger equation for the electronic degrees of freedom (${\bf r}$), resulting in a mere parametric dependency of the wave function on the nuclear configuration (${\bf R}$) (i.e. a continuous family of wave function), denoted as $\Psi({\bf r} | {\bf R})$.
This enables the application of the variational principle to each nuclear configuration independently, and a 
transferable wave function can be optimized by simultaneously minimizing the Rayleigh quotient for each configuration.

Given a probability density $\rho(\mathbf{R}): \Omega\rightarrow\mathbb{R}_{\geq0}$ over a subspace of molecular configuration space $\Omega\subseteq\mathbb{R}^{3M}$, the cost function $\mathcal L^\text{sp}$ is generalized to be minimized if and only if the trial wave function is, in expectation, simultaneously optimal across all sampled molecular geometries:

\begin{equation}
    \mathcal{L}^\text{tr}\big[\Psi_{\boldsymbol{\theta}}\big] =~\underset{\mathbf{R}\sim \rho(\mathbf R)}{\bigE}~\underset{{\mathbf{r}\sim |\Psi_{\boldsymbol{\theta}}(\mathbf{r}\left|\mathbf{R})\right|^2}}{\bigE}\left[ \frac{\hat{H}(\mathbf R)\Psi_{\boldsymbol{\theta}}(\mathbf{r}|\mathbf{R})}{\Psi_{\boldsymbol{\theta}}(\mathbf{r}|\mathbf{R})}\right]\,,
\end{equation}
where the outer expectation value is taken over some predefined distribution of nuclear configurations $\rho({\bf R})$ (see Section \ref{sec:methods:nuclear_sampling} for more details), while the dependence of the wave function and Hamiltonian on the nuclear configuration $\bf R$ within the inner expectation value indicates that the electronic integrals are computed at a fixed nuclear geometry.
During iterative optimization, the expectation value over nuclear geometries is approximated by batches of configurations, while the expectation value over the electronic degrees of freedom is evaluated as before using batches of electron positions sampled from $p({\bf r}) \sim|\Psi_{\boldsymbol \theta}({\bf r} | {\bf R})|^2$ independently for each nuclear configuration, using MCMC.

The above outlined methodology can be extended to the computation of excited states \cite{entwistle2023,pfauAccurateComputationQuantum2024,szabo2024,schatzle2025}.
The treatment of electronically excited states on an equal footing with the ground state can be achieved through additional penalty terms in the loss function.
In the present work, we employ the methodology described by Sch\"atzle \textit{et al.} \cite{schatzle2025}.
In addition to orthogonality constraints the approach can be used to target specific spin states, via a spin-penalty term in the loss function enabling us to directly compute the triplet state of ethylene without orthogonality constraint to the singlet ground state and the two singlet states of the \ce{H2 + NH} reaction without the overhead of considering other, intermediate spin states.

\subsubsection{Shared ansatz for a molecular dataset}
All results presented here were obtained using the transferable ansatz introduced in earlier work by Sch\"atzle \textit{et al.}
The ansatz is adapted from the Psiformer architecture \cite{glehnSelfAttentionAnsatzAbinitio2022}, which exhibits excellent single-point accuracy and has the following form
\begin{equation}
    \Psi (\mathbf{r}) = e^J \sum_k \det[\phi^k_j ({\bf h}^\text{elec}_i)\varphi^k_j({\bf r}_i) ]_{i,j=1}^N \,
\end{equation}
where $J$ is a permutation invariant Jastrow factor implementing the electronic cusp condition, $\varphi^k_j$ are one-particle exponential envelope functions that ensure the correct decay of the wave function, and $\phi^k_j$ are many-body orbitals constructed from electron embeddings ${\bf h}^\text{elec}_i$.
The many-body electron embeddings are generated through a repeated application of the self-attention mechanism on initial electron features consisting of the relative location to the nuclei and a spin encoding.

To adapt this architecture for a geometrically transferable setting, we facilitate the flow of nuclear geometry information through the ansatz by introducing nuclear embeddings
\begin{equation}
    E_{IJ} = \text{Concat}\big(\text{OneHot}(Z_J), |{\bf R}_I - {\bf R}_J|, {\bf R}_I - {\bf R}_J \big) \, ,
\end{equation}
\begin{equation}
     {\bf h}^{\text{nuc},0}_I =
     \text{MLP}_\text{embed} \left(
     \sum_J^{N_\text{nuc}}\text{MLP}_\text{edge} \left(E_{IJ}\right) \right) \, ,
\end{equation}
with MLP denoting multi layer perceptrons using the SiLU activation function and a single hidden layer, and feeding these nucleus embeddings into the self-attention block of the Psiformer along with the electron representations ${\bf h}^{\text{elec},0}$.
Inspired by the Born-Oppenheimer approximation we direct messages from nuclei to electrons and have no electron information flowing to the nuclei, which also reduces the computational cost of the electronic Laplacian.
The final electron embeddings after $L$ interaction cycles ${\bf h}^{\text{elec}^L}_i$ are projected to the many-body orbitals similar to the original Psiformer, while the final nuclear representations ${\bf h}^{\text{nuc},L}_I$ are used to predict the parameters of the envelope functions.


\subsubsection{Chemically motivated sampling of nuclear geometries}\label{sec:methods:nuclear_sampling}
Previous attempts at obtaining geometrically transferable neural network ansatzes with zero-shot chemical accuracy focused on small, well-defined regions of the nuclear configuration space, e.g. dissociation of diatomic molecules or low dimensional cuts across larger potential energy surfaces \cite{scherbela2022,gao2022}.
In this setting it is feasible to define the distribution of nuclear configurations simply by selecting a number of representative geometries \cite{scherbela2022}, and optionally applying small Gaussian \cite{gao2022,scherbela2024} or vibrational normal mode based noise \cite{scherbela2023}.
However, as the number of nuclear degrees of freedom increases and more complex geometric changes must be described, enumerating all representative geometries becomes prohibitive.
To overcome this limitation,  we introduce the concept of Z-matrix based nuclear geometry sampling, with which a more systematic and fine-grained control over the sampled geometries can be achieved.

By working exclusively in internal coordinates the sampled geometries are guaranteed to respect physical symmetries, while defining the structure in terms of bond lengths, angles, and torsion angles, high energy structures are more straightforward to avoid than in cartesian coordinates.

All distributions of nuclear configurations used for VMC training in this work are therefore defined via the distributions of the distance, angle, and dihedral values appearing in the system's Z-matrix or closely related internal coordinate representations.
These, potentially correlated, distributions can be chosen appropriately for the concrete problem at hand, i.e. to yield geometries only around structures suspected to be relevant for the molecule's chemistry, while avoiding atom clashes and other high energy structures.
The concrete definition of nuclear geometry distributions are discussed separately for each demonstrated application in Section \ref{sec:results}.

\subsubsection{Electron warping and equilibration}
When sampling electron positions from the probability density associated with the square of the wave function $p({\bf r} | {\bf R})\sim |\Psi_{\boldsymbol{\theta}}({\bf r} | {\bf R})|^2$ via MCMC, it is of utmost importance that the MCMC chains are well equilibrated.
In practice, it is therefore computationally advantageous to keep in memory the equilibrated MCMC chains from the previous training iteration and perform a small number of re-equilibrating steps to account for changes in the distribution due to the update of the wave function parameters as opposed to equilibrating freshly initialized MCMC chains.
In the transferable setting, however, subsequent iterations may use different molecular configurations and electron positions saved from the previous iteration might end up with extremely low probability from which re-equilibration might be costly or even infeasible.
To alleviate this issue, upon an update of the molecular configuration, electrons are warped to follow the nuclei \cite{umrigar1989}
\begin{equation}
   \mathbf R \rightarrow \mathbf R'  \Rightarrow \mathbf{r}'_i = \mathbf{r}_i + \sum_I f(|\mathbf r_i - \mathbf R_I|) (\mathbf R'_I - \mathbf R_I)\,,
\end{equation}
where $\mathbf R'$ and $\mathbf r'$ are the new and $\mathbf R$ and $\mathbf r$ are the old nuclei and electron positions respectively, and $f$ is a warp function, i.e. a nearest neighbor assignment \cite{gao2022a}. 
This ensures that after a change in the nuclear positions, the minimum distance to a nucleus does not increase for any of the electrons, and in practice greatly reduces the electron re-equilibration steps necessary after a nuclear update. 

\subsection{Nuclear forces}\label{sec:methods:nuclear_forces}
After discussing the techniques used in obtaining converged deep-learning solutions for large, continuous sections of high dimensional PESs, in this section we elaborate on the algorithms that might be used to extract nuclear forces from the obtained wave functions.

\subsubsection{Monte Carlo estimators}
Nuclear forces are traditionally obtained from VMC wave functions via the Hellmann--Feynman theorem, which states that for variationally optimal wave functions
\begin{equation}
    \frac{\text{d} E}{\text{d}\lambda} = \braket{\Psi | \frac{\text{d}\hat H}{\text{d}\lambda} | \Psi} \, ,
\end{equation}
where $\lambda$ is a continuous parameter, the nuclear positions in our case.
However, the expectation value derived this way does not have the zero variance property of the energy itself \cite{assaraf2003}.
To remedy this issue, along with those of a potential bias in the forces due to the wave function not being at variational optimum, Assaraf and Caffarel have introduced a series of improved estimators with which both variance and bias can be reduced \cite{assaraf2003}.
The downside of these estimators is significantly increased per iteration computational cost, compared to the bare estimator.

In the present work we systematically compare the accuracy and efficiency of a number of traditional nuclear force estimators (see Sec.~A.2 of the Supplementary Information).
On top of the bare force estimator ($F_\text{bare}$) we evaluate a zero-variance estimator \cite{assaraf1999}
\begin{equation}
    F_\text{AC-ZV} = F_\text{bare} + \frac{(\hat H - E_\text{loc})\frac{\text{d} \Psi}{\text{d} \lambda}}{\Psi} \, ,
    \label{eqn:ac-zv}
\end{equation}
where $E_\text{loc}$ is the local energy of the wave function, i.e. $\frac{\hat H \Psi}{\Psi}$.
The second term of \eqref{eqn:ac-zv} ensures that it exhibits a similar zero-variance property as the energy \cite{assaraf2003}.
On the other hand, the evaluation of this term requires computing $\hat H\frac{\text{d} \Psi}{\text{d} \lambda}$, necessitating the evaluation of third derivatives of the ansatz, a costly task even with modern automatic differentiation tools.
This cost can be greatly reduced by replacing the exact wavefunction derivative with a simplistic expression $\frac{\text{d} \Psi}{\text{d}\lambda} \approx Q \Psi$.
Assuming that the derivative is taken with respect to the $x$ coordinate of nucleus $I$ ($X_I$), $Q$ takes the form
\begin{equation}
    Q = Z_I \sum_{i=1}\frac{x_i - X_I}{r_{iI}} \, , 
\end{equation}
with $x_i$ denoting the $x$ coordinate of the $i$th electron.
This approximation reproduces the correct asymptotic behaviour of the wave function around the nuclei, where most of the variance in the force is originating.
The obtained reduced-cost variant of the $F_\text{AC-ZV}$ estimator is denoted $F_\text{AC-ZVQ}$.

On top of its unfavourable variance, the bare force estimator might also suffer from a systematic bias, as the Hellmann--Feynman theorem assumes a variationally optimal trial wave function, which is only approximately fulfilled in practice.
This bias can be reduced, by adding an additional term to the improved estimators
\begin{equation}
    F_\text{AC-ZVZB} = F_\text{AC-ZV} + 2 (E_\text{loc} - \braket{E_\text{loc}}) \frac{\frac{\text{d} \Psi}{\text{d}\lambda}}{\Psi} \, ,
\end{equation}
and
\begin{equation}
    F_\text{AC-ZVQZB} = F_\text{AC-ZVQ} + 2 (E_\text{loc} - \braket{E_\text{loc}}) \frac{\frac{\text{d} \Psi}{\text{d}\lambda}}{\Psi} \,,
\end{equation}
with the latter being a mix of the minimal (Q) zero variance estimator and the exact zero bias contribution of \citet{assaraf2003}, that we identify as computationally optimal for our purpose.
Note that computing the new term does not necessitate the evaluation of third derivatives, and can be done with negligible extra cost over the local energies themselves.

\subsubsection{Gaussian process regression}
\label{sec:methods:energy_surface_fitting}

While the force estimators outlined above could be used on their own to perform geometry optimization, additional improvements can be made by taking further advantage of the energy data that automatically becomes available when computing QMC forces, and by re-using QMC data across similar geometries encountered during geometry optimization.
Both of these can be accomplished by constructing local models of the PES, fitted to QMC energy and force data. We employ Gaussian process regression (GPR), which has been used successfully to approximate complex PESs of small molecules \cite{deringer2021} and natively handles uncertainty arising due to the stochasticity of the Monte Carlo method. 

The method allows us to obtain a local approximation of the PES at a cost similar to a single-point energy or force evaluation (see Sec.~A.2 of the Supplementary Information), while providing robust uncertainty estimates, and enabling the re-use of the generated data for subsequent, similar geometries.

Instead of evaluating the energy and force solely at the geometry of interest $\mathbf{R}$, we sample small Gaussian perturbations of the nuclear positions to obtain a range of geometries $\{\mathbf{R}_i\}_{i=1}^n$ in its neighborhood.
For the sampled geometries energies and forces are extracted from the optimized transferable wave function $\Psi_{\boldsymbol\theta^*}$ by evaluating the respective expectation values with Monte Carlo estimation
\begin{align}
\begin{split}
    &\bar{E}_i = \left\langle \hat H(\mathbf{R}_i)\right\rangle_{\Psi_{\boldsymbol\theta^*}(\mathbf{r}|\mathbf{R}_i)} \approx E(\mathbf{R}_i)\\
    &\bar{\mathbf{F}}_i = \left\langle \left.\frac{\mathrm{d}\hat H}{\mathrm{d} \mathbf{R}}\right|_{\mathbf{R}_i} \right\rangle
    _{\Psi_{\boldsymbol\theta^*}(\mathbf{r}|\mathbf{R}_i)} \approx \mathbf{F}(\mathbf{R}_i) \,,
\end{split}
\end{align}
where for the latter the force estimators of Sec.~\ref{sec:methods:nuclear_forces} are employed.
The data is assembled into a dataset $\mathcal{D}$ of molecular configurations $ \mathbf{X} = [\mathbf{R}_1, \ldots, \mathbf{R}_n]^\mathrm{T}\in\mathbb{R}^{n\times3M}$ with noisy energy and force labels $\mathbf{Y} = [\bar{E_1}, \bar{\mathbf{F}}_1, \ldots, \bar{E}_n, \bar{\mathbf{F}}_n]^\mathrm{T}\in\mathbb{R}^{n\times(1+3M)}$. 
The number of molecular configurations $n$, the sampling radius and the level of stochastic sampling noise controlled through the number of sampled electron configurations per molecule are hyperparameters of the optimization that slightly affect the overall efficiency of the method. The total number of samples per force evaluation is adjusted automatically to meet accuracy thresholds for the GPR estimates.

To fit a local approximation of the PES we define a Gaussian process (GP)
\begin{equation}
    \tilde{E}(\cdot)\sim\mathcal{GP}(m,k)\,.
\end{equation}
We use a constant prior mean $m$ and employ the radial basis function (RBF) kernel
\begin{equation}
    k_{\mathrm{RBF}}(\mathbf{R},\mathbf{R}')=\sigma_k^2\exp\!\left(-\frac{\|\mathbf{R}-\mathbf{R}'\|^2}{2\ell_k^2}\right)\,,
\end{equation}
with the two tunable parameters $\sigma_k$ and $\ell_k$ determining the kernel amplitude and correlation lengths scale.
The hyperparameters of the kernel as well as the intrinsic noise level are chosen to maximize the marginal (log) likelihood of the data under the model
\ifarxiv
\begin{align}
\begin{split} 
\log L = -\frac{1}{2}\big(\mathbf{y}^T(&\mathbf{K}+\sigma_n^2\mathbf{I})^{-1}\mathbf{y}\\
&+\log\det(\mathbf{K}+\sigma_n^2\mathbf{I})+n\log 2\pi\big)\,.
\end{split}
\end{align}
\fi
While GPR for fitting large molecular datasets usually employs atomic descriptors \cite{deringer2021}, in our case it is sufficient to directly operate on the internal coordinates of the molecule.
Conditioning the GP on the dataset yields a posterior process
\begin{equation}
    \tilde{E} \mid \mathcal{D} \sim \mathcal{GP}(m^*,k^*)\,,
\end{equation}
which implies Gaussian-distributed finite-dimensional marginals. In particular, for any input
\(\mathbf{R}\),
\begin{equation}
    \tilde{E}(\mathbf{R}) \mid \mathcal{D}
    \sim
    \mathcal{N}\!\bigl(m^*(\mathbf{R}),k^*(\mathbf{R},\mathbf{R})\bigr)\,.
\end{equation}
Since Gaussian priors and likelihoods are conjugate, the posterior mean $m^*(\mathbf{R})$ and (co)variance $k^*(\mathbf{R},\mathbf{R})$ can be obtained in closed form 
\begin{align}
\begin{split}
    &m^*(\mathbf{R})=m + k\big(\mathbf{R},\mathbf{X}\big)(\mathbf{K}+\sigma_n^2\mathbf{I})^{-1}(\mathbf{Y}-m)\\
    &k^*(\mathbf{R},\mathbf{R}) = k(\mathbf{R},\mathbf{R}) - k(\mathbf{R},\mathbf{X})(\mathbf{K}+\sigma_n^2\mathbf{I})^{-1}k(\mathbf{X},\mathbf{R})\,,
\end{split}
\end{align}
where $\sigma_n$ models the intrinsic noise of the data and $\mathbf{K}$ is the augmented kernel matrix, that contains contributions for the energy and forces, respectively
\ifarxiv
\begin{align}
\begin{split}
    \mathbf{K}&= \begin{bmatrix} \mathbf{K}_{EE}~~~~\mathbf{K}_{EF}\\\mathbf{K}_{FE}~~~~\mathbf{K}_{FF}\end{bmatrix}\in\mathbb{R}^{n\times(1+3M)\times n\times(1+3M)}\\
    \mathbf{K}_{EE} &=\begin{bmatrix}k(\mathbf{R}_i,\mathbf{R}_j)\end{bmatrix}_{i,j=1}^n\\&= \begin{bmatrix}k(\mathbf{R}_1,\mathbf{R}_1)~ \cdots~ k(\mathbf{R}_1,\mathbf{R}_n)\vspace{-4pt}\\ 
    \vspace{-1pt}\vdots~~~~~~~~~~~\ddots~~~~~~~~~~~\vdots\\
    k(\mathbf{R}_n,\mathbf{R}_1)~ \cdots~ k(\mathbf{R}_n,\mathbf{R}_n)\\ 
    \end{bmatrix}\\
    \mathbf{K}_{EF} &= \begin{bmatrix}\nabla_{\mathbf{R}_j} k(\mathbf{R}_i,\mathbf{R}_j)\end{bmatrix}_{i,j=1}^n\\
    \mathbf{K}_{FF} &= \begin{bmatrix}\nabla_{\mathbf{R}_i}\nabla_{\mathbf{R}_j}^\top k(\mathbf{R}_i,\mathbf{R}_j)\end{bmatrix}_{i,j=1}^n\,.
\end{split}
\end{align}
\fi
The prediction for the energy at geometry of interest $\mathbf{R}$ is then identified as the posterior mean, i.e. the expectation value of the posterior distribution, with an uncertainty estimate given by the posterior variance.
For sufficiently differentiable kernels the derivative of a GP is again a GP and can be computed analytically.
In practice, we employ automatic differentiation to obtain the gradient and Hessian of the GP.

\subsection{Characterization of the PES}\label{sec:methods:geomopt}
This section briefly describes the methods employed to find equilibrium structures, transition states and entire minimum energy pathways (MEPs) having access to forces and Hessians extracted from the transferable wave function.
\subsubsection{Geometry optimization}
For a detailed review of the techniques used in molecular geometry optimization and related tasks we refer to the work of Schlegel \cite{schlegel2011}, while here we summarize only the most important aspects.
Since computing the gradient using the GPR approach outlined in Section \ref{sec:methods:energy_surface_fitting} also provides an estimate of the Hessian at no additional cost, here we exclusively use exact second-order optimization methods.
Given an initial nuclear geometry ${\bf R}_0$, a step towards the minimum energy geometry can be taken via the (regularized) Newton update formula
\begin{equation}
    \delta {\bf R} = - (H + \lambda I)^{-1}{\bf g} \,,
    \label{eqn:newton_step}
\end{equation}
where $\mathbf{g}$ is the energy gradient with respect to the nuclear configuration, $H$ is the corresponding Hessian and $\lambda I$ is a diagonal regularization term to stabilize the inversion.
To prevent erroneous, arbitrarily large steps, the step size is restricted via a simple trust region approach, that is the step computed above is scaled by the factor
\begin{equation}
    \delta\tilde{\mathbf{R}} = \alpha\delta\mathbf{R}\,,~\text{where}~~\alpha=\begin{cases}
        1 \;\; \text{if} \;\; |\delta {\bf R}| < \tau \\
        \frac{\tau}{|\delta {\bf R} |} \;\; \text{otherwise}~~.
    \end{cases}
\end{equation}
The trust radius $\tau$ is updated after every iteration based on how well the quadratic PES model predicted the change in energy.
Given the ratio $\rho = \frac{E_\text{new} - E_\text{old}}{\delta{\bf R}^T{\bf g} + \frac{1}{2} \delta {\bf R}^T H \delta {\bf R}}$, where $E_\text{old}$ and $E_\text{new}$ denote the energy before and after the last step, the trust radius is updated via
\begin{equation}
    \tau' = \gamma\tau\,,~\text{where}~~\gamma=\begin{cases}
        \frac{2}{3} \;\; \text{if} \;\; \rho < 0.6 \\
        \frac{6}{5} \;\; \text{if} \;\; 0.6 < \rho < 1.4 \\
        1 \;\; \text{otherwise}~~~~~~~~~~~.
    \end{cases}
\end{equation}
To search for transition states, i.e. saddle points where exactly one eigenvalue of the Hessian is negative, an image function is minimized \cite{helgaker1991}, by flipping the sign of the negative Hessian eigenvalue before performing the Newton step of \eqref{eqn:newton_step}.
To perform constrained geometry optimizations, e.g. in the context of a minimum energy pathway optimization, the reduced-restricted method of Anglada is used \cite{anglada1997}.

\subsubsection{Minimum energy path optimization}\label{sec:methods:mepopt}
While simple standalone geometry optimizations are suitable to determine e.g. the equilibrium geometry of a molecule, characterizing reaction pathways necessitate more complex workflows such as MEP optimizations. 
An MEP is a string of nuclear geometries passing through the transition state and connecting the configuration of the reactants with that of the products. 
To perform this task with transferable deep-learning VMC, we implement a simple variant of the double-ended growing string method \cite{zimmerman2013}.
In this approach, one starts with two guess structures which are optimized to obtain the reactant and product geometries.
From these two geometries, a string of nodes connecting them is iteratively built by adding one node at a time alternately to the ``reactant end" and the ``product end" of the string.
A node is added by first interpolating towards the other end of the string, using a fixed step size that results in a uniform distribution of nodes along the string.
Then, the position of the newly added node is optimized in the coordinates orthogonal to the local tangent direction of the string, using constrained geometry optimization.
Once the string of nodes has connected, the node with the highest energy is selected as initial guess for the transition state search.
After the transition state is found, the strings connecting it to the reactants and products are re-grown independently from each other, to obtain an overall uniform distribution of nodes, and to reduce the chance of sharp bends appearing in the final MEP.


\ifarxiv
\paragraph{Code availability}
\fi
The code with which the DeepQMC simulations were run will be made available on GitHub (https://github.com/deepqmc/deepqmc) under the MIT license, upon publication of the manuscript.

\ifarxiv
\paragraph{Acknowledgements}
\fi
This work was funded by the Deutsche Forschungsgemeinschaft (DFG, German Research Foundation) under Germany’s Excellence Strategy -- The Berlin Mathematics Research Center MATH+ (EXC-2046/1, project ID: 390685689) projects (AA1-6, AA2-8) and Deutsche Forschungsgemeinschaft (DFG, German Research Foundation) project NO825/3-2. The computing time granted by the Resource Allocation Board and provided on the supercomputer Lise at NHR@ZIB as part of the NHR infrastructure is gratefully acknowledged. The calculations presented here were conducted with computing resources under the project bec00266.

\ifarxiv
\paragraph{Author contributions}
\fi
ZS, PBS and FN conceived the project.
ZS and PBS developed the method in full detail.
ZS and PBS wrote the computer code.
The numerical experiments were conceived, carried out, and analyzed by ZS and PBS.
ZS and PBS wrote the manuscript with input from FN.
FN supervised the project.
Funding was acquired by FN.

\appendix
\bibliographystyle{unsrtnat}
\bibliography{transferability}
\clearpage
\onecolumn
\section{Supplementary analysis of methodological considerations for ab-initio geometry optimization with deep-learning variational Monte Carlo}
In this section, additional experiments are included supporting the findings of the main manuscript.
In Sec. \ref{sec:results:n2} continuous and grid-based sampling of molecular configurations are compared during variational Monte Carlo (VMC) training.
Then, in Sec. \ref{sec:results:estimator-comparison}, the use of the Assaraf--Caffarel approximate zero variance zero bias estimator to generating force data for our Gaussian process regression (GPR) models is validated.

\subsection{Comparison of continuous and grid-based sampling during VMC training}\label{sec:results:n2}
\begin{figure}[h!]
    \centering
    \includegraphics[width=\linewidth]{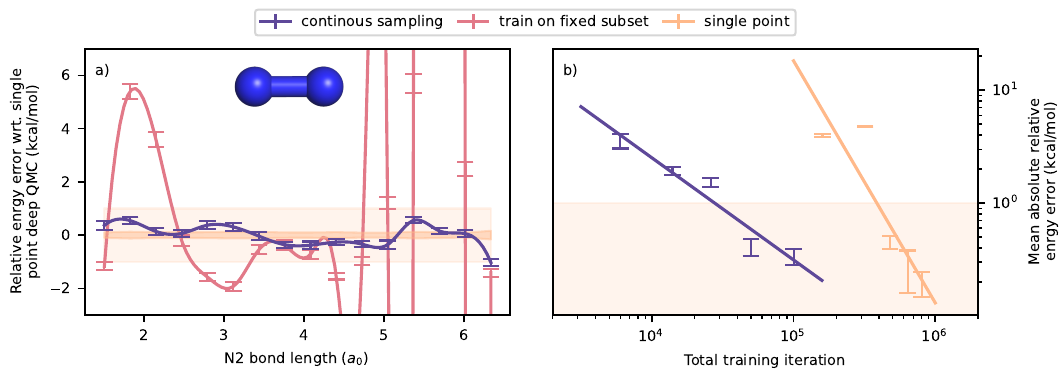}
    \caption{{\bf Continous transferable training converges to single point results.} 
    a) Relative energy error from converged single point optimizations for 16 geometries of the N2 molecule. Continuous sampling and training on the test set converge to chemical accuracy wrt. to the reference on all 16 geometries. Training on a sparse subset of four geometries (1.5\,bohr, 2.93\,bohr, 4.36\,bohr, 6.5\,bohr) results in large interpolation errors.
    b) Mean absolute error in relative energies wrt. converged single point simulations along the entire dissociation curve throughout the optimization. Each data point is an evaluation of the wave function(s) on all geometries at the respective training iteration.
    }
    \label{fig:n2_dissociation}
\end{figure}
To demonstrate the benefits of continuous transferable optimization, we consider the task of simulating the nitrogen dimer dissociation.
We compare continuous transferable training, with bond lengths sampled between 1--9 bohr during training, with transferable training on a grid of four fixed geometries using the same overall number of electron samples. 
To evaluate the transferable simulations we assemble a test set of sixteen evenly spaced molecular configurations along the bond dissociation, for which we run independent single-point optimizations as reference data.
Firstly, we focus on the accuracy of the transferable wave functions, when evaluated on the test set.
While training on the sparse dataset shows significant energy deviations upon interpolation, continuous transferable optimization yields energies well within chemical accuracy of extended single point simulations across the entire dissociation curve (Fig. \ref{fig:n2_dissociation}\,a)). 
This indicates that dense coverage of the configuration space is needed for good interpolation and renders the wave function trained on the fixed dataset unsuitable for geometry optimization.
With the sparsity of the training dataset only ever increasing with the dimensionality of the PES, training on fixed grids quickly becomes an unworkable solution for a consistent approximation of large PESs.
Next we look at the convergence rate of the simulations with the total number of training iterations, comparing the mean absolute error of relative energies with respect to the single point reference along the sixteen geometries of the test set (Fig. \ref{fig:n2_dissociation}\,b)).
Continuous transferable optimization achieves chemical accuracy using an order of magnitude less computational resources compared to single point simulation (note that this speedup heavily depends on the number of geometries being considered). 
This brings the overall cost of simulating the dissociation curve close to that of a VMC optimization for a single molecular geometry using the ``traditional'' deep QMC setup.

Being able to obtain a consistent transferable wave function across relevant subsections of molecular configuration space at a cost comparable to that of a single-point simulation constitutes the foundation for the ab-initio geometry optimization with deep QMC.

\subsection{Comparison of various nuclear force estimators}
\label{sec:results:estimator-comparison}
\begin{figure}[h!]
    \centering
    \includegraphics[width=\linewidth]{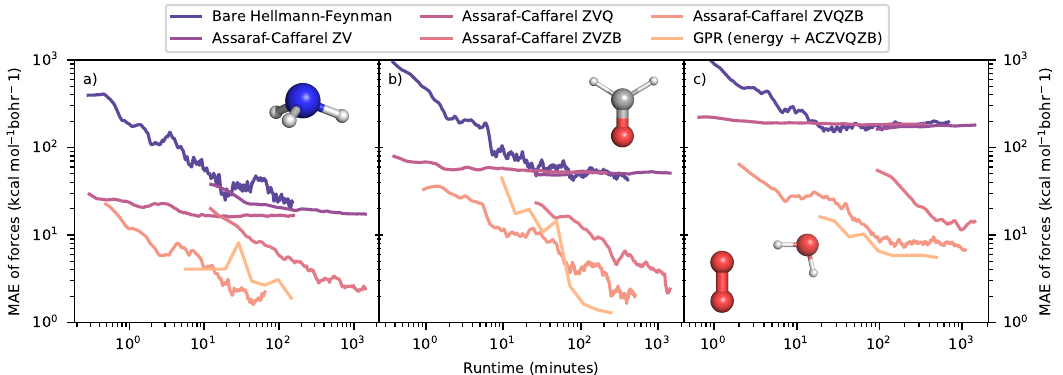}
    \caption{\textbf{Benchmarking the cost-accuracy ratio of various VMC nuclear force estimators.} The comparison is carried out on ammonia, formaldehyde, and O\textsubscript{2} + H\textsubscript{2}O molecular geometries. Finite difference forces obtained with the CCSD(T) based W1 protocol are used as reference. The mean absolute deviations are computed across all nuclei and all cartesian directions of each geometry. The wall clock runtimes were measured using four Nvidia A100 GPUs in parallel.}
    \label{fig:force-estimator-comparison}
\end{figure}
As discussed in Section 1.1 of the manuscript, our GPR models are fitted on both energy and force data extracted from the QMC wave function.
However, while obtaining zero-variance, zero-bias estimates of the energy is straightforward, this is not the case for nuclear forces, for which a number of advanced QMC estimators have been derived by Assaraf and Caffarel \cite{assaraf1999,assaraf2003}.
In this section, we compare the performance of several of these force estimators on three systems ranging from 10--26 electrons, to determine the most advantageous estimator to use for generating force data for the fitting of GPR.
The bare Hellmann--Feynman, Assaraf--Caffarel zero variance (ACZV), zero-variance zero-bias (ACZVZB) estimators, as well as their approximated variants (ACZVQ and ACZVQZB, respectively) \cite{assaraf1999,assaraf2003} are compared, on the task of computing nuclear forces for a single molecular geometry.
Unlike previous force estimator benchmarks in the context of deep QMC \cite{qian2022}, which have primarily focused on sheer accuracy, here we are interested in the practically more relevant cost-accuracy ratios.
Accordingly, on Figure \ref{fig:force-estimator-comparison}, achieved force errors are plotted against the computational cost of obtaining them.
The considered ammonia, formaldehyde, and O\textsubscript{2} + H\textsubscript{2}O geometries are all weakly correlated, therefore reference forces obtained with the CCSD(T) based composite W1 method are used.
The vertical axes of Figure \ref{fig:force-estimator-comparison} show the mean absolute error (MAE) of the nuclear forces across all nuclei and all cartesian directions.
It is clear that without the zero bias correction, the bare Hellmann--Feynman, ACZV, and ACZVQ estimators can't converge to acceptable error levels, regardless of the number of samples considered.
Moving to the two zero-bias corrected estimators, it is evident that the approximated ACZVQZB variant is superior, mainly due to it's greatly reduced computation cost, achieving the same MAE as much as one order of magnitude faster, when compared to the full ACZVZB estimator.
Finally, the performance of the GPR estimator fitted on energies and ACZVQZB forces is also shown on \ref{fig:force-estimator-comparison}, and it is found to be as efficient as the best performing single-point ACZVQZB estimator.
On the formaldehyde and O\textsubscript{2} + H\textsubscript{2}O geometries, it converges even faster than the underlying ACZVQZB estimator, which we attribute to the fact that the GPR estimator is fitted on energies in addition to forces, which, due to their own zero-variance property, form valuable additional information, and are available at no additional cost over the ACZVQZB forces.

In conclusion, among the traditional force estimators, the ACZVQZB estimator is deemed most appropriate in practice, as it beneficially combines the zero-bias correction with a low-cost yet accurate approximate version of the zero-variance correction.
Furthermore, it is verified that even in a setting where only a single geometry is considered, the GPR estimator performs at least as well as the underlying traditional force estimator, and therefore we can expect a strict improvement over single-point estimators when moving to a many-geometry setting, where data generated for previous, nearby geometries can be reused.
The GPR estimator fitted on energy and ACZVQZB force data is therefore used in all further experiments.
\vfill
\section{Description of the procedures used to obtain the more complex training geometry distributions}
\FloatBarrier
\subsection{The C2H4 triplet relaxation}
For the relaxation of the C2H4 triplet state we follow the setup of Barborini and coworkers \cite{barboriniStructuralOptimizationQuantum2012} and optimize on a four-dimensional internal-coordinate space comprising the C–C and C–H distances, the H–C–H angle, and the H–C–C–H torsion, which are constrained to be identical for all occurrences.
For the optimization of the wave function we used four independent uniform distributions around the ground state equilibrium geometry with widths reported in Table \ref{tab:c2h4_geom}.

\begin{table}[ht]
    \centering
    \caption{The widths (or min -- max if unsymmetrical) of the uniform distributions used to define the training geometry distribution for C2H4. The atoms are not numbered, since all occurrences  of the internal coordinates that are considered as bonds in the equilibrium geometry are equal.}
    \begin{tabular}{p{7cm}|cc}
        Description & Value & Unit \\ \hline
        C--C distance & -0.5 -- 1.0 & bohr \\
        C--H distance (4x) & 0.5 & bohr \\
        H--C--H angle (2x) & 0.5 & rad \\
        H--C--C--H dihedral & $\pi$& rad \\
    \end{tabular}
    \label{tab:c2h4_geom}
\end{table}

\subsection{The H$_2$ + NH reaction}
Wu and coworkers \cite{wu2020} provide 25 planar geometries along the investigated reaction, optimized at the CASPT2/6-311G* level of theory.
These geometries are taken as centers of a mixture of uniform distributions with equal weight, defined on the bond lengths and angels of the system (only planar geometries are considered).
The widths of these distributions are selected so that the mixture of 25 distribution forms a roughly continuous, uniform distribution.
The width values are collected in Table \ref{tab:hnh2_geom}.
\begin{table}[ht]
    \centering
    \caption{The widths of the 25 uniform distributions used to define the training geometry distribution for the H$_2$ + NH reaction.
    The hydrogen atoms of the attacking hydrogen molecule are designed H1 (farther from NH) and H2 (closer to NH), while the hydrogen of NH is denoted H3.}
    \begin{tabular}{p{7cm}|cc}
        Description & Value & Unit \\ \hline
        H1 -- H2 distance & 0.755 & bohr \\
        H2 -- N distance & 0.516 & bohr \\
        N -- H3 distance & 0.2 & bohr \\
        H2 -- N -- H3 angle & 1.1 & rad \\
    \end{tabular}
    \label{tab:hnh2_geom}
\end{table}

\subsection{The O$_2$H + OH reaction}
Liu \textit{et al.} \cite{liuAnomalousKineticsReaction2019} provide four geometries along the reaction in question, optimized via a permutation-invariant polynomial neural network interatomic potential, fitted on CCSD(T)-F12 data.
These geometries correspond to the reactant, product, transition state structures, as well as to a local minimum found just before the transition state (CP1).
We interpolate consequtive pairs of these geometries using internal coordinates, to obtain a total of 32 geometries, distributed roughly uniformly along the reaction pathway.
These are then considered the 32 centers of a mixture of uniform distributions with equal weight.
The uniform distributions are defined on the bond lengths, angles, and dihedrals of the system.
Note that while all reference geometries are planar, here we don't enforce planarity in the training dataset, to enable learning of a full dimensional PES, with 9 degrees of freedom.
Just as for the H$_2$ + NH reaction, the widths collected in Table \ref{tab:o2hoh_geom} are selected such that the resulting mixture distribution is roughly continuous, and approximately uniform.
\begin{table}[ht]
    \centering
    \caption{The widths of the 32 uniform distributions used to define the training geometry distribution for the O$_2$H + OH reaction.
    The atoms of the O--O--H molecule are designated O1, O2, and H1, respectively, while the atoms of the OH molecule are denoted O3 and H2.}
    \begin{tabular}{p{7cm}|cc}
        Description & Value & Unit \\ \hline
        O1 -- O2 distance & 1.0 & bohr \\
        O2 -- H1 distance & 0.12 & bohr \\
        O1 -- O2 -- H1 angle & 1.4 & rad \\
        H1 -- O3 distance & 0.22 & bohr \\
        O2 -- H1 -- O3 angle & 0.7 & rad \\
        O1 -- O2 -- H1 -- O3 dihedral & 1.516 & rad \\
        O3 -- H2 distance & 0.6 & bohr \\
        H1 -- O3 -- H2 angle & 0.6 & rad \\
        O1 -- H1 -- O3 -- H2 dihedral & 1.57 & rad \\
    \end{tabular}
    \label{tab:o2hoh_geom}
\end{table}
\FloatBarrier
\section{Employed hyperparameters for the geometry optimizations and MEP searches}

\begin{table}[ht]
    \centering
    \caption{Collection of hyperparameters employed during the geometry optimizations and minimum energy pathway searches. `data radius' refers to the maximum Euclidian distance in between the target geometry and the sampled data points considered for fitting the local Gaussian approximate potential.}
    \begin{tabular}{c|ccccc}
        Hyperparameter & formaldehyde & ammonia & ethylene & H$_2$ + NH & O$_2$H + OH \\ \hline
        initial trust radius (bohr) & 0.2   & 0.2     & 0.05      & 0.2        & 0.25 \\
        Newton-step regularizer ($\lambda$) & $10^-2$ & 0 & 0 & 0 & $10^{-2}$ \\
        data radius (bohr) &  0.2     &   $\infty$    &   $\infty$   &      0.2   &  0.2     \\
        energy convergence criterion (Hartree) & $10^{-3}$ & $5\times10^{-4}$ & $5\times10^{-4}$ & $10^{-3}$ & $10^{-3}$ \\
    \end{tabular}
    \label{tab:geomopt_params}
\end{table}
\end{document}